\documentclass[english,aps,amssymb,prx,twocolumn]{revtex4}
\pdfoutput=1
\usepackage[T1]{fontenc}
\usepackage[latin9]{inputenc}
\usepackage{amsmath}
\usepackage{graphicx}
\usepackage{amssymb}
\usepackage{esint}
\usepackage{tensor}
\usepackage{xfrac}
\usepackage{float}
\makeatletter
\@ifundefined{textcolor}{}
{
 \definecolor{WHITE}{gray}{1}
 \definecolor{RED}{rgb}{1,0,0}
 \definecolor{GREEN}{rgb}{0,1,0}
 \definecolor{BLUE}{rgb}{0,0,1}
 \definecolor{CYAN}{cmyk}{1,0,0,0}
 \definecolor{MAGENTA}{cmyk}{0,1,0,0}
 \definecolor{YELLOW}{cmyk}{0,0,1,0}
 }

	\usepackage{subfigure}
	\usepackage{array}
	\usepackage{xcolor}

\renewcommand{\phi}{\varphi}
\renewcommand{\epsilon}{\varepsilon}

\usepackage{epsfig}
\makeatother

\usepackage{babel}

\begin{document}

\title {Topological phase transitions in glassy quantum matter}
\author{Isac Sahlberg$^{1}$}
\author{Alex Weststr\"om$^{2,3}$}
\author{Kim P\"oyh\"onen$^{2,4}$}
\author{Teemu Ojanen$^{1,2}$}
\email[Correspondence to ]{teemu.ojanen@tuni.fi}
\affiliation{$^{1}$Computational Physics Laboratory, Tampere University, P.O. Box 692, FI-33014 Tampere, Finland}
\affiliation{$^{2}$Department of Applied Physics (LTL), Aalto University, P.~O.~Box 15100,
FI-00076 AALTO, Finland }
\affiliation{$^{3}$Westlake University, Hangzhou, Zhejiang Province, China  }
\affiliation{$^{4}$QuTech and Kavli Institute of Nanoscience, Delft University of Technology, 2600 GA Delft, The Netherlands }

\date{\today}
\begin{abstract}
{\bf Amorphous systems have rapidly gained promise as novel platforms for topological matter. In this work we establish a scaling theory of amorphous topological phase transitions driven by the density of lattice points in two dimensions. By carrying out a finite-size scaling analysis of topological invariants averaged over discrete and continuum random geometries, we discover unique critical properties of Chern and $\mathbb{Z}_2$ glass transitions. Even for short-range hopping models the Chern glass phase may persist down to the fundamental lower bound given by the classical percolation threshold. While the topological indices accurately satisfy the postulated one-parameter scaling, they do not generally flow to the closest integer value in the thermodynamic limit. Furthermore, the value of the critical exponent describing the diverging localization length varies continuously along the phase boundary and is not fixed by the symmetry class of the Hamiltonian. We conclude that the critical behaviour of  amorphous topological  systems exhibit characteristic features not
observed in disordered systems, motivating a wealth of new research directions. }
\end{abstract}

\maketitle
\bigskip{}


{\bf Introduction--} While topological classification of matter is in principle completely independent on the symmetry-breaking classification, a vast recent literature is devoted to the increasingly subtle interplay of topology and spatial order \cite{chiu}. However,  by adopting a completely  complementary starting point, a number of recent studies have identified amorphous systems \emph{without} reference to a band structure as fruitful platforms for topological states \cite{shenoy,mitchell,xiao,mansha,bourne,chern,agarwala}. This crucial property sets amorphous topological systems  apart from disordered and Anderson topological insulators \cite{li,groth}  where non-trivial topology relies on residual spatial order.  Amorphous topological states are extremely appealing for two reasons. In contrast to crystalline topological states, they do not rely on the specific spatial distribution of their microscopic constituents, thus being exceptionally robust. Furthermore, the possibility to fabricate topological states through randomly located dopants could allow accessing a whole new class of designer topological systems. This aspect was recently highlighted in concrete proposals for amorphous topological superconductors \cite{poyh2} and insulators \cite{minarelli}. The existence of amorphous topological states has become a well-established fact with a rapidly growing number of novel proposals.   

\begin{figure}[h]
\includegraphics[width=0.99\linewidth]{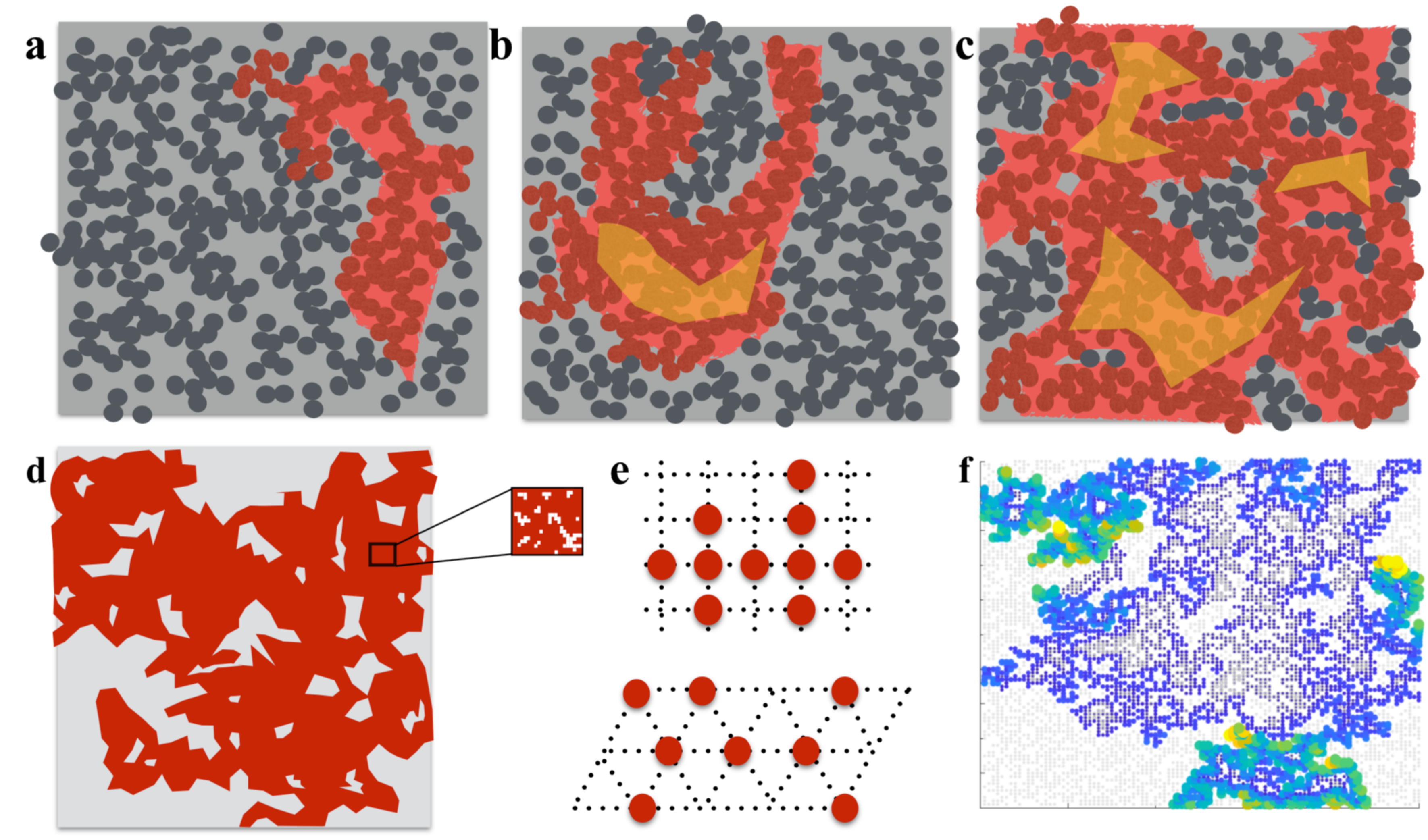}
\caption{ \textbf{a}-\textbf{c}  illustrate the proliferation of topologically non-trivial wave functions through amorphous systems as the density of lattice sites is increased. The dots represent randomly distributed dopants that support electronic orbitals hybridizing with nearby sites. The red and orange regions represent the magnitude of wave functions with the largest spatial extent. Configuration \textbf{c} supports an extended state, thus defining a topologically non-trivial system.  \textbf{d}: The course-grained geometry of large clusters of discrete percolation lattices is largely insensitive to the lattice-scale details shown in the inset. The square and triangular site percolation configurations shown in \textbf{e} give rise to statistically similar long-distance random geometry. \textbf{f}: The colored area represents the modulus of topological wavefunction  $|\psi|$ on a critical percolating cluster on a square lattice.  }
\label{fig:fsurface}
\end{figure}
Previously it has been shown that various hopping models with randomly distributed dopants will undergo a topological phase transition at sufficiently high density \cite{shenoy, poyh2}. The purpose of this work is to establish a quantitative description of this phase transition in representative topological models with randomly-generated geometries in two dimensions. Our work highlights the interplay of physics of topological matter \cite{volovik,bernevig1}, the Anderson localization \cite{evers,huck} and classical percolation theory \cite{stauffer}. In contrast to semiclassical or effective network studies \cite{kramer}, our theory is based on fully quantum-mechanical evaluation of topological invariants of representative microscopic models averaged over random geometries and studying their finite-size scaling properties in the vicinity of the critical density.

A topological phase transition in an amorphous system can be thought of as a proliferation of wave functions carrying topological indices through the random lattice as depicted in Fig.~1 a-c.  We consider both continuum and discrete random geometries that are commonly studied in classical percolation theory. Discrete lattices with randomly occupied sites represent systems where random dopants have preferred absorption sites. Furthermore, when the correlation length of the underlying classical percolation problem is comparable or larger than the system size, discrete and continuum random lattices describe a similar large-scale geometry insensitive to short-range details, as illustrated in Fig.~1 d-f. We establish that the divergence of the localization length of the wave functions carrying topological quantum numbers satisfy the scaling behaviour $\xi=(p-p_c^Q)^{-\bar{\nu}}$,  where $p$ denotes the probability of a lattice site to be occupied, $p_c^Q$ is the critical probability and $\bar{\nu}>0$ is a critical exponent. Continuum models are shown to obey a similar relation where the site occupation probability is replaced by density of particles $\rho$. Unexpectedly, the non-trivial phase for paradigmatic short range hopping models may survive down to the fundamental lower limit given by the classical site percolation threshold. Equally unexpectedly, the critical exponent $\bar{\nu}$ is non-universal and varies continuously within a symmetry class of the Hamiltonian. This behaviour is in sharp contrast to excessively studied topological transitions in disordered systems. Our findings stimulate a large number of research directions in the rapidly rising field of amorphous topological matter.
 

{\bf Theoretical approach--} The purpose of our work is to formulate a quantitative theory of topological phase transitions in amorphous 2d systems. We consider systems with broken time-reversal symmetry characterized by non-zero Chern numbers as well as time-reversal invariant systems classified by a $\mathbb{Z}_2$ invariant \cite{schnyder}. The minimal representative lattice models for these cases have two and four orbitals per lattice site. The minimal models are expected to capture a generic behaviour of the topological phase transition because amorphous systems do not exhibit spatial symmetries that would enforce additional band degeneracies.  An effective tight-binding Hamiltonian for the Chern insulator can be expressed as 
\begin{equation}
H_1=\begin{pmatrix}\label{h1}
(2-M)\delta_{ij}+T_{ij} & iT_{ij}e^{-i\phi_{ij}}  \\
iT_{ij}e^{i\phi_{ij}} 	& -(2-M)\delta_{ij}-T_{ij} 
\end{pmatrix},
\end{equation}
where $M$ is the onsite energy difference of the two orbitals in the units of characteristic hopping amplitude and $T_{ij}=-\frac12 e^{-r_{ij}/\eta}\theta(R-r_{ij})$ describes the spatial decay of the hopping amplitudes. Here $r_{ij}=|\mathbf{r}_i-\mathbf{r}_j|$ is the distance between sites $i,j$, the parameters $\eta, R$ describe the decay of hopping, and the phase factor is given by $e^{i\phi_{ij}}=\frac{r_{ij}^x+ir_{ij}^y}{r_{ij}}$, where $r_{ij}^x = x_i-x_j$.
Introducing two additional orbitals, the studied time-reversal invariant $\mathbb{Z}_2$ model can expressed as
\begin{equation}\label{h2}
H_2=\begin{pmatrix}
H_1 & \Gamma \\
\Gamma^\dagger & H_1^*
\end{pmatrix},
\end{equation}
and describes two time-reversed copies of $H_1$ coupled by the off-diagonal block $\Gamma=i\alpha\,\mathrm{Diag}\left[iT_{ij}e^{-i\phi_{ij}}, iT_{ij}e^{i\phi_{ij}}\right]$. This model can be thought of as an amorphous Bernevig-Hughes-Zhang (BHZ) model \cite{bernevig2} with a Rashba-type inversion-breaking term parametrized by $\alpha$. If the models are discretized on a square lattice with $\eta=\infty$ and $R$ just above the lattice constant, these models reduce to a paradigmatic two-band Chern insulator $H_1(k)=\sin k_x \sigma_x + \sin k_y \sigma_y+\left[2-M-\cos k_x -\cos k_y \right]\sigma_z$  and an inversion-broken BHZ model $H_2(k)=\mathrm{Diag}\left[ H_1(k),H_1(-k)^*\right]+\alpha \left[-\sin k_x \tau_y +\sin k_y \tau_x\sigma_z \right]$, where $\sigma_i$ and $\tau_i$ are two sets of Pauli matrices. 

\begin{figure*}
\includegraphics[width=0.99\linewidth]{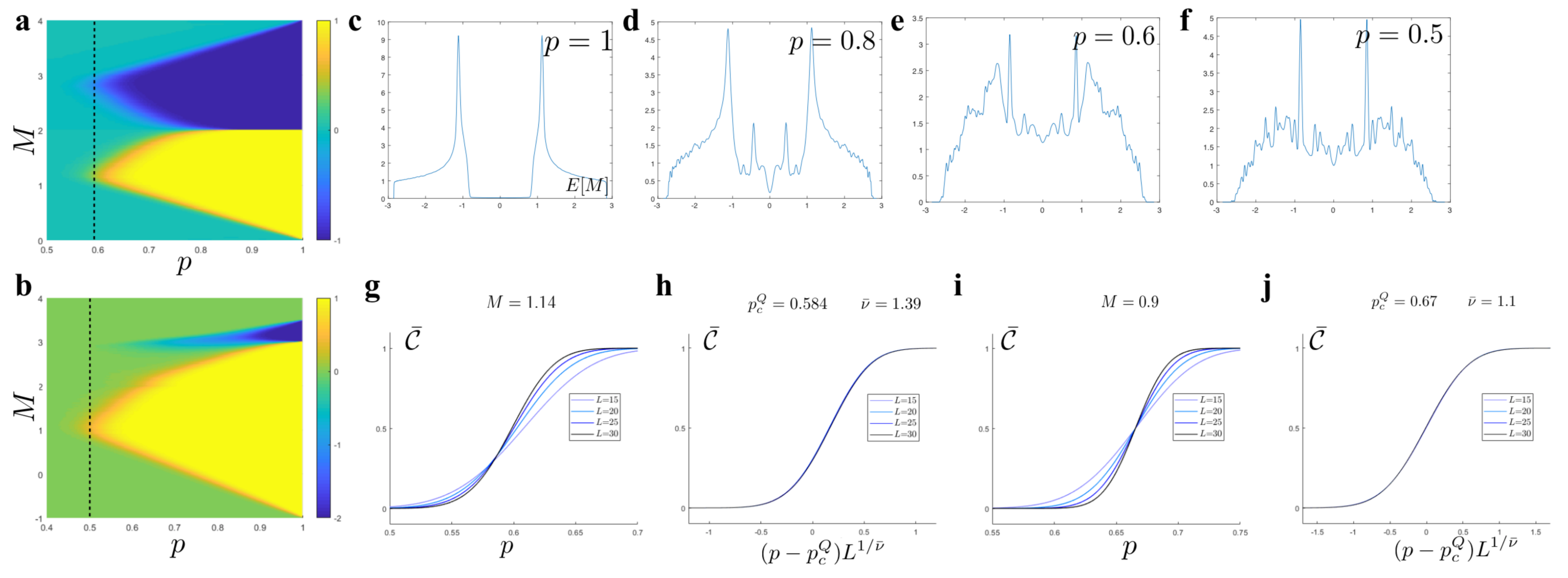}
\caption{ \textbf{a}: Configuration averaged Chern number for a nearest-neigbour Chern insulator generated by random square lattices. The topological index is averaged over 20 configurations on a $20\times20$ lattice. The dashed line marks the classical percolation threshold $p_c^{cl}$ \textbf{b}: the same as \textbf{a} but for triangular random lattices.\textbf{c}-\textbf{f}: Configuration averaged DOS of a square lattice Chern glass at optimal value $M=1.14$ for different fillings $p$. The figures are calculated by Kernel Polynomial Method \cite{weisse} for $80\times80$ lattices using 300 moments, 200 random vectors and averaged over 200 realizations. \textbf{g}: Configuration averaged Chern number for different system sizes as a function of the site filling probability for square lattice generated geometries at the optimal point $M=1.14$. \textbf{h}: Scaling collapse of \textbf{g}. The extracted $p_c^Q$ and $\bar{\nu}$ values are in agreement with the classical percolation threshold $p_c$ and correlation length exponent $\nu$.  \textbf{i} and \textbf{j}: Same as  \textbf{g} and \textbf{h} but for $M=0.9$.  }\label{fig:chern}
\end{figure*}

We will study the behaviour of the topological invariants of models \eqref{h1} and \eqref{h2} averaged over different random geometries as a function of the density of lattice sites, assuming that half of the orbitals are populated. To gain insight on the thermodynamic limit,  we will study the finite-size scaling behaviour of the topological invariants. The invariants are calculated by the real-space Chern number and Bott index algorithms that require a single diagonalization of the Hamiltonian per configuration \cite{zhang2,loring}. We will consider continuum problems with randomly distributed lattice sites characterized by a density per unit area $\rho$, as well as discrete random geometries familiar from the percolation theory shown in Fig.~1 d-f. Percolation theory studies the cluster properties on different lattices where the sites are independently occupied with probability $p$. A cluster in the discrete problem is defined as a collection of lattice points that are connected by a nearest-neighbour connecting path and in continuum problems as a set of points that belong to a union of site-centred discs of fixed radius.  Above one dimension there generically exists a critical probability $0<p_c^{cl}<1$ or critical density $0<\rho_c^{cl}$  above which infinite systems contain an infinite cluster, defining a second-order phase transition. A course-grained long-distance random geometry generated by different percolation lattices is similar to continuum percolation problems. Indeed, as confirmed here explicitly, the critical properties for discrete and continuum random geometries are essentially similar.  

Our scaling hypothesis for the amorphous topological phase transition is formulated in terms of topological invariants averaged over random geometries. We propose that the averaged Chern number obeys the scaling
\begin{equation}\label{scale}
\bar{\mathcal{C}}=f[(p-p_c^Q)L^{1/\bar{\nu}}]
\end{equation}
as a function of the occupation probability and linear system size $L$. Here $f(x)$ is an a priori unknown scaling function which approaches to 0 (1) for $x\ll0$ ($x\gg0$).  For continuum problems we postulate a similar expression with $p$ and $p_c^Q$ substituted by particles per unit area $\rho$ (in appropriate units) and its critical value $\rho_c^Q$.  The scaling form \eqref{scale} has important ramifications. First, it implies that the phase transition is sharp in the thermodynamic limit, taking place at the critical value $p_c^Q$ which we call the topological quantum percolation threshold. Second,  $\bar{\nu}$ is a critical exponent describing the diverging localization length of wave functions carrying the Chern number through the relation $\xi=(p-p_c^Q)^{-\bar{\nu}}$ near the transition. In addition, due to the connection between the Chern number and the Hall conductance, Eq.~(\ref{scale}) also describes the scaling of the Hall response (in the units of $e^2/h$) in amorphous systems. This provides a concrete connection between our theory and observables. The $\mathbb{Z}_2$ invariant can be evaluated through a configuration-averaged spin Bott index $\bar{\mathcal{C}}_s$ \cite{huang}, for which we assume a similar scaling form. It should be stressed that while the scaling hypotheses for topological invariants are superficially similar to the scaling of percolation probability \cite{stauffer}, they characterize the topology of quantum ground states of models \eqref{h1}, \eqref{h2}, which is conceptually completely independent on the classical percolation problem.

{\bf Chern glass transition--}  We first consider the discrete random geometries.  The topological phase diagrams as a function of density for square and triangular random lattices are illustrated in Fig.~2 \textbf{a}-\textbf{b}. The non-trivial phases in amorphous systems are protected by a mobility gap rather than a true energy gap. This is reflected in the fact that the gapped spectrum of a pristine system is rendered gapless for $p<1$ by randomly missing sites, as illustrated by the density of states (DOS) in Figs.~1 \textbf{c}-\textbf{f}. The location of the phase boundary between trivial and non-trivial phases and the corresponding critical filling $p_c^Q$ is seen to depend strongly on the value of the mass parameter $M$.  Strikingly, for the optimal values of $M$ the topologically non-trivial state is seen to extend all the way down to the classical percolation threshold $p_c^{cl}$. While there is recent convincing evidence of a localization transition for nontopological quantum percolation models for $p<1$, it is thought to be possible only in the close vicinity of $p=1$ \cite{schubert1, schubert2, dillon}.  Therefore it is remarkable that the topological phase may persist all the way down to the classical percolation threshold.  At the site percolation threshold, at which point there is no longer a meaningful separation between bulk and edge, there appears a critical cluster characterized by a fractal dimension $d=1.9$.  For the optimal values of $M$, the Chern glass model avoids Anderson localization even at $p_c^{cl}$. As illustrated in App.~B of the Supplemental Information (SI), topological wave functions efficiently penetrate the bottlenecks of the critical cluster. It should be noted that  $p_c^{cl}$ gives the lower bound for the existence of the topological phase in the thermodynamic limit since below that the system is disconnected and cannot support extended states for nearest-neighbour models. The signatures of a topological phase below $p_c^{cl}$ reflect finite-size effects and numerical accuracy.     

\begin{figure*}
\includegraphics[width=0.99\linewidth]{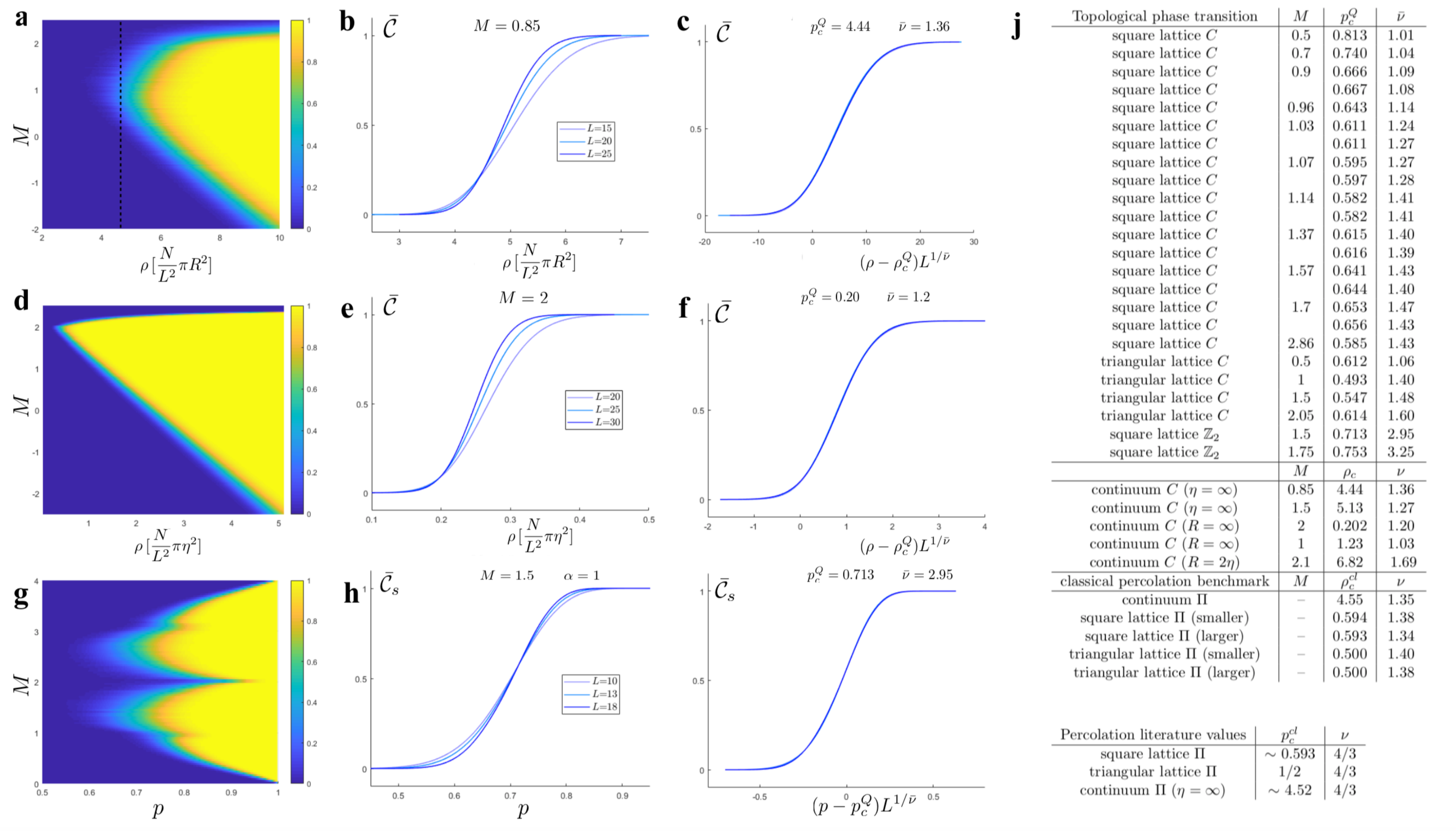}
\caption{\textbf{a}: Configuration averaged Chern number for continuum Chern glass model with constant hopping inside disc with radius $R$. The topological index is averaged over 20 configurations on a $16\times16$ lattice. The dashed line marks the classical percolation threshold $p_c^{cl}=4.52$  \cite{quitanilla}. \textbf{b}: Topological transition curves for the model in \textbf{a} with $M=0.85$. \textbf{c}: Scaling collapse of \textbf{b}, yielding $p_c^{Q}=4.44$ and $\bar{\nu}=1.36$ in agreement with the classical values within numerical accuracy.  \textbf{d}: the same as  \textbf{a} but for exponentially decaying hopping with $R=\infty$.  \textbf{e}:  Topological transition curves for the model in \textbf{d} with $M=2.0$. \textbf{f}: Scaling collapse of  \textbf{e}. \textbf{g}: Configuration averaged $Z_2$ phase diagram on a square lattice with nearest-neighbour hopping and $\alpha=1$. The topological index is averaged over 40 configurations on a $16\times16$ lattice.  \textbf{h}:  Topological transition curves for the model in \textbf{g} with $M=1.5$ and $\alpha=1$. \textbf{i}: Scaling collapse of the curves in \textbf{h}.  \textbf{j}: Table of calculated critical data for amorphous topological phase transitions, performed with sample sizes between 300 and 1000. Two values for the same parameters correspond to two different statistical samples with the same number of realizations. Below we have included our results for known classical percolation problems for similar systems sizes to benchmark our method of calculating the transition curves and extracting scaling data. The benchmark calculations can be compared to the percolation literature values (bottom). }
\label{fig:continuum}
\end{figure*}

The onset of the topological phase transition and its scaling behaviour is illustrated in Fig.~2 \textbf{g}-\textbf{j}. The method of calculating the configuration-averaged transition curves is explained in App.~A in the SI.  For a fixed $M$, the topological index exhibits very accurate finite-size scaling of the postulated form \eqref{scale}, implying that the localization length of the topologically non-trivial states diverges as $\xi_Q=(p-p_c^Q)^{-\bar{\nu}}$ near the critical filling $p_c^Q$. From the scaling collapse we can estimate the values of the critical filling $p_c^Q$ and critical exponent $\bar{\nu}$. While the phase diagram shows that $p_c^Q$ depends sensitively on $M$, the properties of Quantum Hall transitions in disordered systems \cite{huck,kramer} might suggest that the critical exponent $\bar{\nu}$ could be universal in a symmetry class of Hamiltonians \cite{altland}. \emph{However, this is not the case.}  We observe that the value of $\bar{\nu}$ varies even by a factor of 1.6 as a function of  $M$. As discussed in App.~A, the relative accuracy of extracted $p_c^Q$ and $\bar{\nu}$ values is expected to be around $5\%-10\%$. This estimate follows from benchmarking our method of evaluating the transition curves for classical percolation problems with comparable system sizes and statistical sampling and comparing to values in the literature. In topological and classical problems the main source of error comes from the limited configuration sampling and is of the same order of magnitude. Therefore the numerical uncertainty is very small compared to the variation of  $\bar{\nu}$ along the phase boundary and can be ruled out as a source for observed non-universality of  $\bar{\nu}$. For the optimal $M$ values for which the topological threshold and the classical threshold agrees $p_c^Q\approx p_c^{cl}$, the extended non-trivial states are restricted to the critical cluster. Thus the linear extent of these states should be of the order of the diameter of the classical critical cluster and the value of $\bar{\nu}$ should agree with the correlation length exponent of classical percolation $\nu=4/3$.  Indeed, our results reproduce this fact within numerical accuracy. The values of critical data for selected points are given in the table in Fig.~\ref{fig:continuum}. It is a remarkable feature of the Chern glass transition that the value of \emph{a configuration-averaged Chern number is not universal at $p_c^Q$}, in particular it does not need to be a half-integer. This is in a striking contrast to the disordered quantum Hall transitions where the critical value is a half-integer and the Hall conductance flows to the closest integer multiple of $e^2/h$ in the thermodynamic limit \cite{pruisken}. An amorphous system can flow to $e^2/h$ even if the the initial value is, say, $0.3e^2/h$ as implied by Fig.~2 \textbf{g}.

The topological properties of the continuum Chern glass are in qualitative agreement with the discrete models. First, to make a precise comparison with the classical disc percolation, we consider a continuum model without exponential decay  ($\eta=\infty$) but with a constant hopping amplitude within the radius $R$. In Fig.~3 \textbf{a}-\textbf{c} we have plotted the topological phase diagram as a function of density $\rho$ and studied the scaling properties of the transition. We discover again that the non-trivial phase may reach the vicinity of the classical disc percolation threshold $\rho_c^{cl}$. The critical exponent $\bar{\nu}$ describing the diverging localization length through $\xi_Q=(\rho-\rho_c^Q)^{-\bar{\nu}}$, is also found to vary strongly along the phase boundary. However, at the optimal point where $\rho_c^{Q}\approx\rho_c^{cl}$, we again expect it to agree with the classical percolation result $\nu=4/3$ for the same reasons as in the discrete models. This expectation is confirmed within numerical accuracy. In Fig.~3 \textbf{d}-\textbf{f} we have plotted results for a physically more realistic exponentially attenuated hopping where the hard cutoff is taken to infinity $R=\infty$. In practice, the cutoff at $R$ is basically irrelevant when $R>4\eta$. The scaling behaviour in the exponential Chern glass model is the same as for the disc model. While the critical exponent does not exhibit universality, the discrete and continuum Chern glass models share qualitatively similar critical properties.

{\bf $\boldsymbol {\mathbb{Z}_2}$ glass transition--} Now we complement our results regarding the Chern glass transition by considering a $\mathbb{Z}_2$ model \eqref{h2} on a random square lattice with a nearest-neighbour hopping. The important technical difference to the Chern glass case is that now we have to work with four orbitals per lattice site and the evaluation of $\mathbb{Z}_2$ invariant the help of the spin Bott index \cite{huang} requires two diagonalizations per configuration. These facts make the study computationally substantially more expensive compared to the Chern glass case, thus we are restricted to smaller systems. Here we concentrate on the $\alpha\neq 0$ case since for $\alpha=0$  the model trivially reduces to the above studied Chern glass problem (mod 2). The phase diagram of the system is shown in Fig.~3 \textbf{g}, illustrating how the topological threshold of the $\mathbb{Z}_2$ transition is increased by the coupling between the time-reversed Chern blocks. Nevertheless, as shown in Figs.~3 \textbf{g}-\textbf{h}, the $\mathbb{Z}_2$ glass transition exhibits the same form of scaling \eqref{scale} near the critical density with a non-universal exponent $\bar{\nu}$. The uncertainty for the extracted values of $\bar{\nu}$ is much larger than for the Chern glass models due to the system size limitations. For example, we estimate that for the parameters in Fig.~3 \textbf{g} the uncertainty is of the order  $\bar{\nu}=3.5\pm 1$.  

The studied $\mathbb{Z}_2$ glass phase diagram also highlights an essential qualitative difference from a phase diagram of a disordered BHZ model. A BHZ model with finite coupling between the Chern blocks $\alpha\neq 0$ and quenched disorder exhibits a well-documented metallic phase between topological and trivial insulating phases \cite{yamakage}. However, in the density-driven $\mathbb{Z}_2$ glass transition we do not observe signatures of such a metallic intermediate state. A metallic state would give rise to extended states on a fixed interval of $p$ and comprise a departure from the observed scaling form \eqref{scale} which implies that the strip of extended states (transition region) shrinks as the system size is increased. This fact further supports our findings that the topological properties of crystalline systems with quenched disorder are qualitatively different from those of amorphous systems.

 
{\bf Discussion and outlook--} Motivated by the rising interest in amorphous topological states, we introduced a scaling theory of density-driven topological phase transitions in glassy systems. Surprisingly, glassy topological systems support unique critical properties sharply different from extensively studied disordered systems. While we showed that glassy systems accurately obey the postulated finite-size scaling relation, the critical exponent describing the delocalization transition is not universal. Furthermore, the Hall conductance does not necessarily flow to the closest integer multiple of $e^2/h$ in the thermodynamic limit. Remarkably, the non-trivial Chern glass phase even for short-range hopping models may persist down to the ultimate lower bound, the classical percolation threshold, where the geometry of the random lattice becomes fractal. For the parameter values at which the topological and classical percolation thresholds meet, the critical exponent describing the delocalization transition is consistent with the correlation length exponent of 2d percolation. This seems to be the only universal aspect of the phase transition. While a numerical approach cannot provide a definite answer to what happens in the thermodynamic limit, we observe single parameter scaling in the accessible system sizes.  There exists substantial evidence \cite{schubert1, schubert2, dillon} that localization in non-topological quantum percolation models in 2d is significantly distinct from the conventional one-parameter scaling theory of disordered systems \cite{abrahams}. Therefore it is to be expected that also the topological transition and critical properties of amorphous topological models have unique properties not encountered in disordered systems. 

  Our findings open a whole new line of research in amorphous topological systems. To list a few topics, we mention a generalization of our results to higher dimensions, characterizing the fractal nature of topological states and transport properties near the critical density. Our work also provides a basis in generalizing the theory of topological semimetals to amorphous systems \cite{yang} .

\appendix

\widetext

\section{Calculation of configuration-averaged topological transition curves}\label{a:transition}

\subsection{Discrete random geometries}

In this appendix we detail our method of calculating configuration-averaged Chern numbers $\bar{\mathcal{C}}(p)$ as a function of the occupation probability $p$ of a single lattice site.  This procedure applies to $\mathbb{Z}_2$ transitions and classical percolation problems in essentially the same form.  The key point is to first calculate the Chern data $\bar{\mathcal{C}}(n)$ as a function of the number of occupied sites $n$ randomly distributed in the lattice, and afterwards analytically deduce $\bar{\mathcal{C}}(p)$ for all $p$. This procedure has crucial advantages over directly sampling configurations with sites occupied by probability $p$. First of all, $p$, in contrast to $n$, is a continuous variable and fixing appropriate $p$ resolution for scaling studies is not a priori clear. Secondly, as made explicit below, $\bar{\mathcal{C}}(p)$ for a single $p$ value is obtained as a weighted average over $\bar{\mathcal{C}}(n)$ with several different $n$. This turns out to reduce statistical fluctuations in the transition curves and to allow more accurate evaluation of $p_c^Q$ and $\bar{\nu}$. This is especially important in the studied problem where evaluation of topological indices, unlike the percolation probability, requires diagonalization of the Hamiltonian, imposing limitations for practical system sizes. We benchmark our method by applying it to finite-size scaling of classical percolation problems and show that percolation thresholds $p_c$ and correlation length exponents $\nu$ can be extracted in good accuracy within the system sizes employed in our study.

Assume we have acquired knowledge of the quantity $\bar{\mathcal {C}}(n)$ as a function of occupied sites $n$ on a finite lattice of $N=L^2$ sites.  In practice, this is done by averaging the topological index  over different configurations with the same $n$. In total, there are $2^N$ possible configurations with varying amounts of occupied sites $n$, and there are ${N\choose n}$ possible configurations for a fixed $n$. 
We would now like to know $\bar{\mathcal {C}}(p)$ for all $p \in [0,1]$. Given a probability $p$ for each lattice site to be occupied, the probability to get exactly $n$ occupied sites out of $N$ possible sites is given by the binomial distribution $P(n,p)$. 
Thus, the desired quantity $\bar{\mathcal {C}}(p)$ can be calculated using the following prescription: 
\begin{equation}\label{C(p)toC(n)}
\bar{\mathcal {C}}(p) =
\sum_n P(n,p) \bar{\mathcal {C}}(n) =
\sum_n {N\choose n} p^n (1-p)^{N-n} \bar{\mathcal {C}}(n) .
\end{equation}

The validity of the procedure can be straightforwardly tested. The comparison of $\bar{\mathcal {C}}(p)$ calculated by applying \eqref{C(p)toC(n)} shows that the method indeed reproduces the correct $\bar{\mathcal {C}}(p)$ curves, as well as drastically reduces the statistical fluctuations compared to sampling configurations with probability $p$ directly. Since the conversion from $\bar{\mathcal {C}}(n)$ is performed analytically, one can choose a much higher density for the probabilities without the need to perform additional diagonalization. In Fig.~4, we have compared the transition curves produced by either fixing $p$ directly, and beginning with even very course sampling of $\bar{\mathcal {C}}(n)$ and converting to $\bar{\mathcal {C}}(p)$ using \eqref{C(p)toC(n)}. Note that all errors come from sampling from the finite set of configurations for fixed $n$---perfect knowledge of $\bar{\mathcal {C}}(n)$  would yield the \emph{exact} transition curve $\bar{\mathcal {C}}(p)$.
\begin{figure}[h]
\centering
\includegraphics[width=0.99\linewidth]{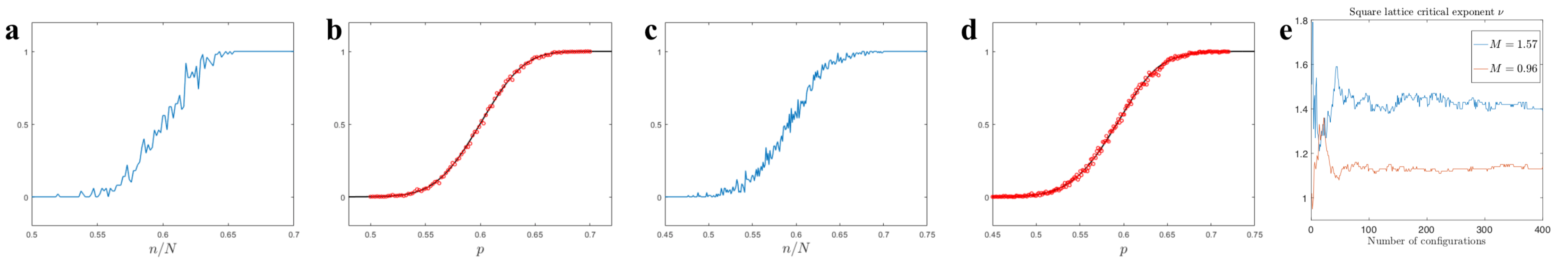}
\caption{The calculation of the transition curves $\bar{\mathcal {C}}(p)$ is substantially more efficient by first calculating the data $\bar{\mathcal {C}}(n)$, and then transforming using Eq.\ \eqref{C(p)toC(n)}, illustrated here. \textbf{a}: A rough sample average of $\bar{\mathcal {C}}(n)$ for $M=1.14$ on a $25\times 25$ square lattice, with a sample size of 50. \textbf{b}: The red dots represent a configuration average of $\bar{\mathcal {C}}(p)$ with a sample size of 1000, while the black graph is the transition curve obtained through the use of \eqref{C(p)toC(n)} on the data in \textbf{a}. In \textbf{c} and \textbf{d}, the same method is illustrated for lattice percolation $\bar{\Pi}(p)$ on a $30\times 30$ square lattice, with the blue graph $\bar{\Pi}(n)$ and the red dots $\Pi(p)$ having a sample size of 100 and 500, respectively. \textbf{e}: The critical exponent calculated by collapsing two $\bar{\mathcal {C}}(p)$ curves for each $M$, as a function of the sample size. Each new set of configurations cumulatively increases the accuracy of the obtained curves, and hence the obtained critical values $p_c^Q, \bar{\nu}$. }
\label{raw_vs_analytical_lattice}
\end{figure}

The method explained above for calculating the transition curves can be applied to $\mathbb{Z}_2$ systems and classical percolation by replacing  $\bar{\mathcal {C}}(n)$  by a spin-Bott index $\bar{\mathcal {C}_s}(n)$, or $\bar{\Pi}(n)$, giving the probability of the existence of a percolating path connecting left and right edges of the sample. Applying \eqref{C(p)toC(n)} gives access to $\bar{\mathcal {C}_s}(p)$ and $\bar{\Pi}(p)$.  To benchmark our method of calculating transition curves, we apply the method to classical percolation problems. In Fig.~3 \textbf{j} in the main text we have collected our critical data. Comparison to the accurate reference values from the literature shows that the relative error in $p_c^{cl}$ and $\nu$ for the comparable system sizes available for topological studies is of the order of 5-10\%. The results obtained for the critical exponents become more accurate as the sample size is increased, as the effects of the statistical sampling decrease, see Fig.~ \ref{raw_vs_analytical_lattice} \textbf{e}. Since the main source of error for classical percolation and topological problems originates from the limited statistical sampling and both exhibit comparable fluctuations, we expect that the error in $p_c^Q, \bar{\nu}$ is of the same order. For classical problems we could go to much larger systems and obtain much more accurate data, but evaluation of topological indices is computationally much more expensive and we are limited to modest system sizes. Nevertheless, the above considerations suggest that we can reach good accuracy also in the topological data.

\subsection{Continuum random geometries}

For continuum models we need to carry out a configuration average of the topological index $\bar{\mathcal {C}}(\rho)$ as a function of density. In contrast to the above, we do not have a fixed underlying lattice where the lattice points are either occupied or unoccupied. 
First of all, we have two separate cases that we will consider: constant hopping in a finite radius $r \le R$, and hopping which decays exponentially $\sim e^{-r/\eta}$. We can denote the cutoff by $\theta (r-R) e^{-r/\eta}$, the former case corresponding to $\eta=\infty$, and the latter to $R=\infty$.
We will first look at the former case; the latter will be handled analogously. Denote by $\rho$ the density $\rho = \frac{\mathrm{Sum\ of\ area\ of\ disks}}{\mathrm{Total\ area}} =  \frac{n \pi R^2}{L^2} = D(n) \pi R^2$, where $n$ is the number of lattice points in our region of size $L^2$, and $D(n)=n/L^2$ is the particle number per system size. For our purposes, in order for $\rho$ to describe a meaningful density, it needs to depend on $R$.
Given a large system with overall particle number density $D(n)$, if we choose a subsystem of size $L^2$, we will, on average, find $n = D(n) L^2$ lattice sites in it. Every lattice point is located either within the confines of the subsystem, or not, completely independently of all other points. What we have described is a Poisson distribution with the intensity $\lambda = D(n) L^2$. The probability to find $k$ lattice points in our system is
\begin{equation}
P_\lambda(k) = \frac{\lambda^k e^{-\lambda}}{k!}, \quad k = 0,1,2,...
\end{equation}
Thus, the quantity $\bar{\mathcal{C}}_R(D)$ for fixed particle number density $D$ and radius $R$ can be calculated using
\begin{equation}\label{C(rho)toC(n)}
\bar{\mathcal{C}}_R(D) =
\sum_n P_{\lambda(D)}(n) \bar{\mathcal{C}}_R(n) ,
\end{equation}
where $\lambda(D) = D L^2$.
Finally, in terms of the density $\rho$, we can give the desired quantity as
$\bar{\mathcal{C}}(\rho) = \bar{\mathcal{C}}(D \pi R^2) = \bar{\mathcal{C}}_R(D)$.
For the case of $R=\infty$, where the exponential decay is controlled by a finite $\eta$, the quantity of interest is instead
$\bar{\mathcal{C}}(\rho) = \bar{\mathcal{C}}(D \pi \eta^2) = \bar{\mathcal{C}}_\eta(D)$.
\begin{figure}[h]
\centering
\includegraphics[width=0.99\linewidth]{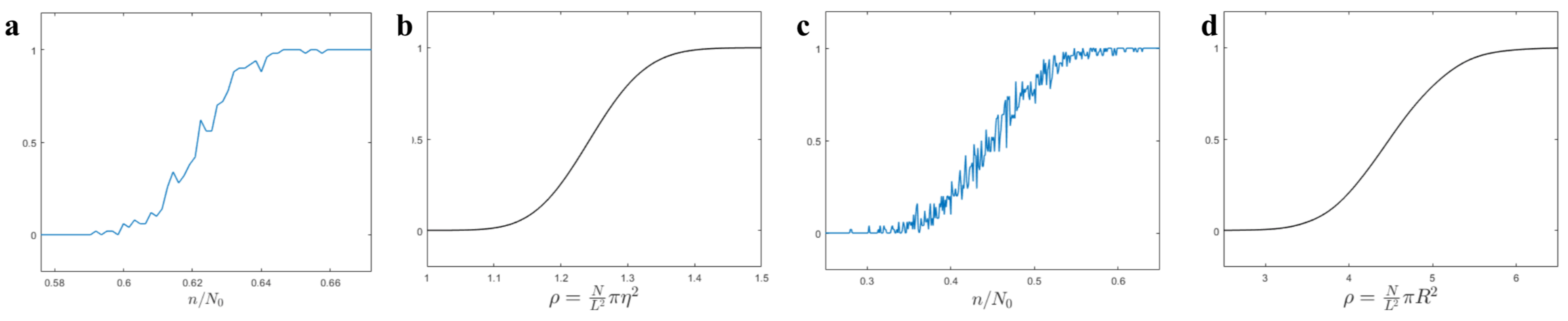}
\caption{Same as in Fig.\ \ref{raw_vs_analytical_lattice}, except for continuous systems with no underlying lattice structure, and where the presription of calculating transition curves is now \eqref{C(rho)toC(n)}. \textbf{a} and \textbf{b} are for the continuum model Chern number $C$ for a 25$\times$25 system with $R=\infty$, and the parameter $M=1$.  \textbf{c} and \textbf{d} are for the inclusive disc percolation for a 30$\times$30 system, where $R$ stands for the maximal hopping radius between lattice points. $N_0 = L^2$ is the number of lattice sites analogous to the maximum $N$ for finite lattices, with the difference that $n$ can here exceed $N_0$.}
\label{raw_vs_analytical_continuum}
\end{figure}
Again, the amount of densities $\rho$ to evaluate $\bar{\mathcal{C}}(\rho)$ can be chosen as high as desired as the conversion from $\bar{\mathcal{C}}(n)$ to $\bar{\mathcal{C}}(\rho)$ is calculated analytically. Note that, in contrast to the binomial distribution for the lattice problem above, the sum $\sum_n$ here goes to infinity, so one must include the terms until $P_{\lambda(D)}(n)$ becomes negligible. Again, the strongest confirmation of the validity of this method comes from benchmarking it for disc percolation. We can calculate the disc percolation probability $\bar{\Pi}(\rho)$ from Eq.~\eqref{C(rho)toC(n)} by replacing  the topological index $\bar{\mathcal{C}}(n)$ by configuration-averaged percolation probability as function of lattice points $\bar{\Pi}(n)$. This enables us to extract  the critical density  and the critical exponent and to compare them to their known values $\rho_c=4.52$ \cite{quitanilla} and  $\nu=4/3$. For system sizes up to $L=30$ with $N=L^2$ particles averaged over 1000 configurations we obtain the values $\rho_c=4.55$, $\nu=1.35$. Since the main source of error for the topological problem and  disc percolation is the limited statistical sampling, we expect similar relative error for $\rho_c^{Q}$ and $\bar{\nu}$.

\section{Critical wave functions at the percolation threshold}

\begin{figure}[h]
\includegraphics[width=0.99\linewidth]{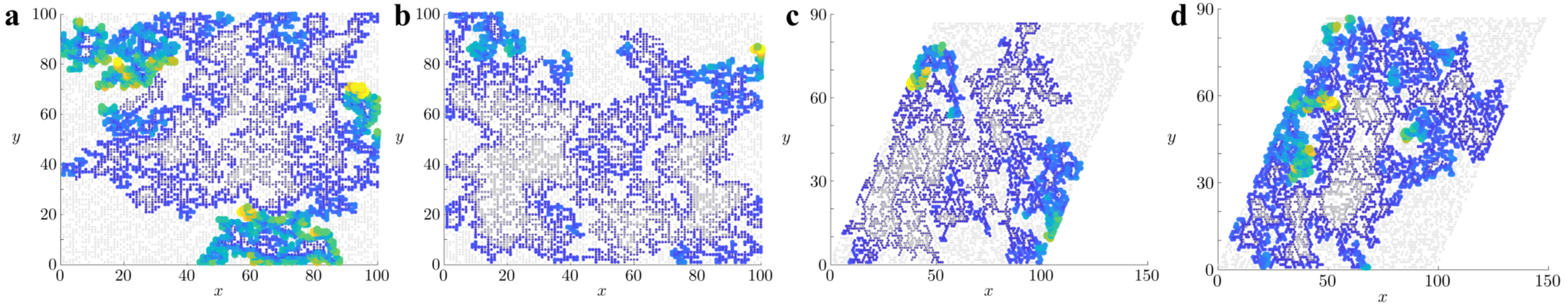}
\caption{ Modulus of critical wave functions $|\Psi|$ on percolating clusters for the nearest-neigbor Chern glass model on $100\times100$ lattices at the percolation threshold. All states have energies $E\lesssim10^{-4}M$.  \textbf{a}-\textbf{b}: Square lattice percolations geometries with $M=1.14$ and $p=p_c=0.59$.  \textbf{c}-\textbf{d}:  Triangular lattice percolation geometries with $M=1$ and $p=0.52$, almost at the threshold $p_c=0.5$ }\label{fig:wavefunctions}
\end{figure}

In the main text we pointed out that for optimal values of the mass parameter $M$, the topologically non-trivial phase may persist down to the site (disc) percolation threshold for short-range hopping models with discrete (continuum) random geometries. The percolation threshold provides a theoretical lower bound for the existence of a non-trivial topological phase for these models since no extended states can exist below the threshold when the underlying system consists of a collection of finite clusters. At the threshold, the underlying critical cluster is a random fractal with dimension d=1.9 in two dimensions \cite{stauffer}. Remarkably, as required by the presence of a non-trivial phase, extended states may survive even in this extreme limit. In the Fig.~\ref{fig:wavefunctions} below we have plotted the modulus of the wave functions living on critical clusters of square and triangular lattices. These extended states nucleate at $E=0$ at the critical density $p_c$ and the linear extent of these states is limited by the critical cluster. This provides a natural explanation for why the localization length exponent at the percolation threshold coincides with the correlation length exponent of the percolation cluster. 
\vspace{-0.1cm}

\end{document}